\documentclass[twocolumn,amsmath,amssymb,prl,superscriptaddress]{revtex4}
\usepackage{graphicx}

\begin{document}

\title{Deterministic Coherent Writing of a Long-Lived Semiconductor Spin Qubit  Using One Ultrafast Optical Pulse}

\author{I. Schwartz}
\affiliation{The Physics Department and the Solid State Institute,
Technion--Israel Institute of Technology, 32000 Haifa, Israel}
\author{D. Cogan}
\affiliation{The Physics Department and the Solid State Institute,
Technion--Israel Institute of Technology, 32000 Haifa, Israel}
\author{E. R. Schmidgall}
\affiliation{The Physics Department and the Solid State Institute,
Technion--Israel Institute of Technology, 32000 Haifa, Israel}
\author{L. Gantz}
\affiliation{The Physics Department and the Solid State Institute,
Technion--Israel Institute of Technology, 32000 Haifa, Israel}
\author{Y. Don}
\affiliation{The Physics Department and the Solid State Institute,
Technion--Israel Institute of Technology, 32000 Haifa, Israel}
\author{M. Zieli\'{n}ski}
\affiliation{Institute of Physics, Faculty of Physics, Astronomy and Informatics, Nicolaus Copernicus University, ul
Grudziadzka 5, PL-87-100 Toru\'{n}, Poland}
\author{D. Gershoni}
\affiliation{The Physics Department and the Solid State Institute,
Technion--Israel Institute of Technology, 32000 Haifa, Israel}
\email[]{dg@physics.technion.ac.il}
\date{\today}

\begin{abstract}
We use one single, few-picosecond-long, variably polarized laser pulse to deterministically write any selected spin state of a quantum dot confined dark exciton whose life and coherence time are six and five orders of magnitude longer than the laser pulse duration, respectively. The pulse is tuned to an absorption resonance of an excited dark exciton state, which acquires non-negligible oscillator strength due to residual mixing with bright exciton states. We obtain a high fidelity one-to-one mapping from any point on the Poincar\'{e} sphere of the pulse polarization to a corresponding point on the Bloch sphere of the spin of the deterministically photogenerated dark exciton.

\end{abstract}

\pacs{}

\maketitle

Future technologies based on quantum information processing (QIP)~\cite{knill2001,monroe2002,ladd2010,imamoglu1999,gisin2002,kimble2008} require the ability to coherently control matter two-level systems, or qubits. Since they are ultrafast and require no contacts, optical means of control are preferred. Spins of charge carriers in semiconductors are promising matter qubits since they complement contemporary leading technologies, specifically those of light sources and detectors.
Semiconductor QDs isolate single carriers and can be easily incorporated into nanophotonic devices, thereby providing an excellent interface between single spins and single photons.
For these reasons, semiconductor quantum dots (QDs) have been the subject of many recent works, which demonstrated significant progress in optical writing, readout, and control of  confined spins ~\cite{berezovsky2008,press2008, greilich2009, gerardot2008, degreve2011,godden2012, poem2011,benny2011prl, kodriano2012, schwartz2015}.

In semiconductors, the absorption of a photon results in the promotion of an electron from the full valence band, across the forbidden band-gap, to the empty conduction band, leaving its spin unaltered. The missing valence band electron (or ``hole") and the conduction band electron form an electron-hole-pair with opposite spin directions, or a bright exciton (BE). The BE forms an integer spin (total spin 1) qubit in the matter.

In a previous work, we demonstrated that, in strain-induced self-assembled quantum dots, the polarization of a resonantly tuned, single picosecond optical pulse can be used to deterministically {\it write} the bright exciton spin qubit in any desired coherent state \cite{benny2011prl}. Such a process is not possible for single spins, where a few pulses are required to prepare the spin in an eigenstate and then a three step (Ramsey) rotation is need
ed to write the spin state at will~\cite{berezovsky2008, press2008, gerardot2008}, a process that can take a few nanoseconds. Moreover, while full coherent {\it control} (``rotation") of the bright exciton can be achieved by one single optical pulse \cite{kodriano2012, muller2013}, single spins require two pulses and free precession in between \cite{berezovsky2008, press2008,greilich2009,godden2012}. However, these advantages of the bright exciton are not very useful, since its lifetime is rather short (sub ns), limited by radiative recombination of the electron-hole pair.

Since light barely interacts with the electronic spin, an electron-hole pair with parallel spin directions is almost optically inactive. Such a pair is called a dark exciton (DE). The DE is also a spin integer (total spin 2) qubit \cite{poem2010}, but since it is optically inactive, its lifetime can be orders of magnitude longer than that of the BE \cite{mcfarlane2009, smolenski2013, schwartz2015, smolenski2015}. Very recently, we have demonstrated that the DE lifetime ($\sim$ 1 $\mu$s) is radiative and that it maintains coherence over at least $\sim$ 100 ns \cite{schwartz2015}. As such, the DE has clear advantages over the bright exciton, provided that it can be externally accessed, despite its optical inactivity.

Here, we exploit the optical activity of an unpolarized absorption resonance to an excited DE state to deterministically photogenerate the DE and write its spin state in any desired coherent superposition of its two eigenstates with high fidelity, using a few picoseconds long polarized optical pulse.
The duration of the pulse is several orders of magnitude faster than both the DE lifetime and coherence time.
This achievement makes the DE an excellent matter qubit. It improves this qubit's usefulness in comparison to a recent publication in which the deterministic generation of the DE and its control was first demonstrated~\cite{schwartz2015}. In Ref~\cite{schwartz2015}, the DE was photogenerated by resonant excitation to its ground state. The weak optical activity (and narrow linewidth) of the DE ground state, however, required pulses of a few tens of nanoseconds duration, and the transition polarization selection rules permitted generation in only one of the DE spin eigenstates. Therefore, full spin initialization required, in addition, three step Ramsey control.

\begin{figure}[b]
\begin{center}
\includegraphics[width=\columnwidth]{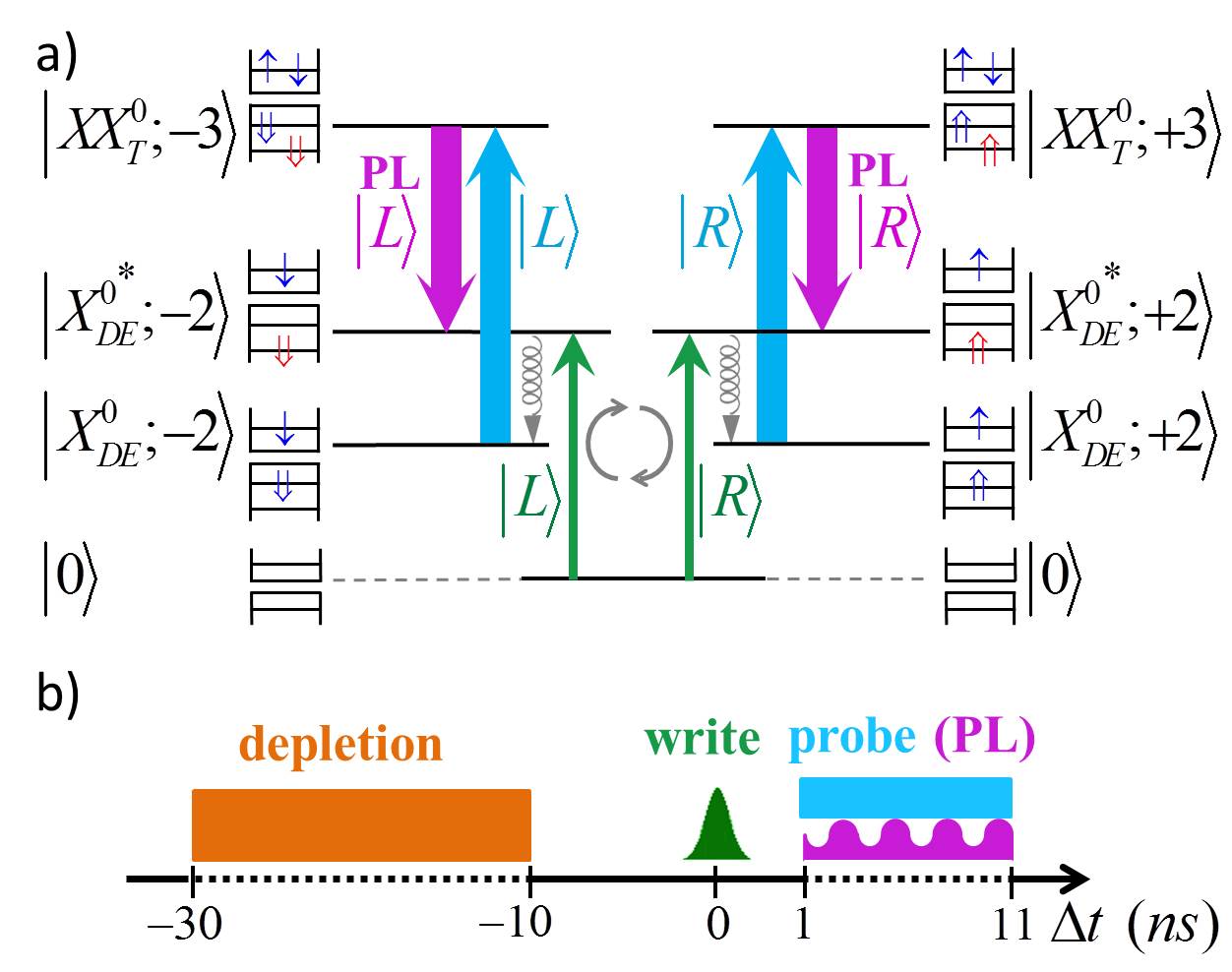}
\caption{Schematic description of the experiment. (a) The relevant energy levels, spin wavefunctions, and optical transitions involved in the experiment. $\uparrow (\Downarrow)$ represents a spin up (down) electron (heavy hole). The blue (red) color represents carrier in the first (second) energy level.
Vertical upward (downward) arrows denote resonant optical excitation (emission) and curly arrows represent non-radiative spin-preserving heavy hole transitions. $|L\rangle$ ($|R\rangle$) describes a left (right) hand circularly polarized optical transition and $|J(=0, \pm 1, \pm 2, \pm 3)\rangle$ describes the total angular momentum (orbital and spin) projection on the QD symmetry axis (growth direction).
A polarized $s$-$p$ resonant laser pulse (green color) excites the DE via its mixing with BE states. The pulse polarization defines the phase and amplitude between the right hand circular polarization induced transition (left side) and the left hand one (right side). The photogenerated DE is therefore a coherent superposition of both transitions, and precesses between the $|+2\rangle$ and $|-2\rangle$ spin states, as indicated by the grey cyclical arrow.
The long circularly polarized probe pulse (teal color) transfers the DE population into $XX^{0}_{T\pm 3}$ biexciton population.
The efficiency of the population transfer depends on the temporal spin polarization of the precessing DE and it is monitored by detecting the emitted photon which results from the $XX^{0}_{T\pm 3}$ biexciton radiative decay (magenta color).
(b) The temporal sequence of optical pulses used in the experiment. A depletion pulse (orange) emptied the QD. The $s$-$p$ resonant write pulse (green) deterministically photogenerated a DE in a coherent superposition of eigenstates. The probe pulse (teal) transferred the DE population to the $XX^{0}_{T\pm3}$ biexciton. The polarization of the emitted PL (magenta) indicates the DE spin state.
}
\label{fig:Scheme}
\end{center}
\end{figure}

Fig. \ref{fig:Scheme} schematically describes the experiment. The DE is deterministically generated and its state is ``written" using a variably polarized short pulse tuned to the DE $s$-like electron(ground energy level $(1e^1)_{\pm 1/2}$ ) - $p$-like heavy hole (first excited energy level $(2h^1)_{\pm 3/2}$) absorption resonance, as shown by the green upward arrows in Fig. \ref{fig:Scheme}(a). The optical writing of the DE spin is made possible due to mixing between the $s$-$p$ DE states and BE states. This mixing is induced mainly by the QD asymmetry~\cite{don2015} and compositional disorder~\cite{zungerprb2009}.
Thus, the overall wavefunctions of these excited DE ($X^{0*}_{DE}$) states contain spin antiparallel states in addition to the spin-parallel states:
$\rm |X^{0*}_{DE};\pm 2 \rangle=\alpha[(1e^1)_{\pm 1/2}(2h^1)_{\pm 3/2}]+\beta[(1e^1)_{\mp1/2}(2h^1)_{ \pm 3/2}]+\gamma[(1e^1)_{\pm 1/2}(2h^1)_{ \mp 3/2}]$f
  where $(1e^1)_{\pm 1/2}$ [$(2h^1)_{\pm 3/2}$] represents a single electron [hole] in its ground [first excited] state with spin projection $\pm 1/2$ [$\pm 3/2$] on the optical axis (notation after Ref. \cite{benny2011prb}), and $|\alpha|>>|\beta|,|\gamma|$. It is this small mixing with spin anti-parallel bright states that provides the optical activity of the excited DE states.

  The excited DE then rapidly relaxes to its ground state ($X^{0}_{DE}$) non-radiatively by a spin preserving \cite{woods2004,heiss2007} phonon emission  (grey curly arrow in Fig. \ref{fig:Scheme}(a)), which is much faster than the radiative recombination rate of these excited states (DE and BE alike~\cite{kodriano2010}). After relaxation to the ground state, the coherent DE state evolves in time. The spin state and population of the DE are probed using excitation to spin-blockaded biexciton states ($XX^{0}_{T\pm3}$) where the two hole spin projections are parallel, as described in Fig. \ref{fig:Scheme}(a). The time evolution of the DE spin state is probed by the absorption of a second polarized, 10 ns long pulse, tuned to the DE$-XX^{0}_{T\pm3}$ biexciton resonance (teal arrows in Fig. \ref{fig:Scheme}(a)). The magnitude of the absorption depends on the DE population and polarization at the photon absorption time. The temporal dependence of the absorption is monitored by detecting the emitted $XX^{0}_{T\pm3}$ biexciton photon (magenta arrows in Fig. \ref{fig:Scheme}(a)). The DE is then optically depleted from the QD~\cite{Schmidgall2015} and the experiment repeats itself from the beginning. The pulse sequence for the experiment is shown in Fig. \ref{fig:Scheme}(b).

\begin{figure}[b]
\begin{center}
\includegraphics[width=\columnwidth]{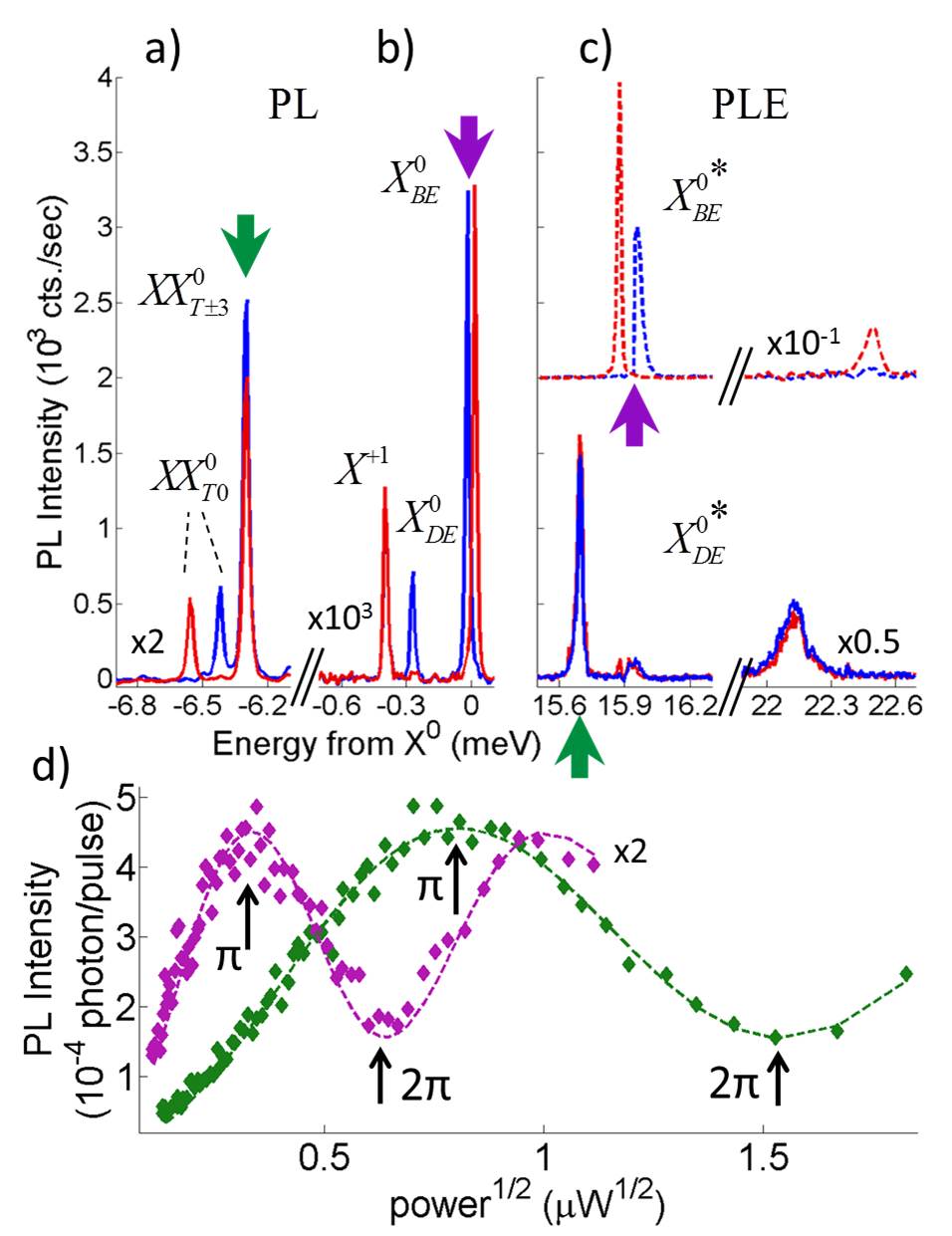}
\caption{(a) and (b) Rectilinear polarization-sensitive PL spectra of the QD. The biexciton (a) and exciton (b) transitions were excited by 5$\mu$W and 10nW 445nm cw laser light, respectively. The observed spectral lines are identified by their initial states. (c) Rectilinear polarization sensitive PL excitation (PLE) spectra of the BE (dashed line) and DE (solid line). Both spectra were obtained simultaneously by varying the energy of one laser while tuning an additional laser to the $XX^{0}_{T_{\pm3}}$ resonance (blue upward arrow in Fig. \ref{fig:Scheme}(a)). The DE PLE was obtained by monitoring the PL emission from the $XX^{0}_{T_{\pm3}}$ line (downward green arrow in (a)) and that of the BE by monitoring the $X^{0}_{BE}$ line directly (downward magenta arrow in (b)). (d) Rabi oscillations of the $X^{0}_{DE}$ (green, measured using the $XX^{0}_{T\pm3}$ emission as a probe) and $X^{0}_{BE}$ (magenta). Color matched downward arrows in (a-b) indicate the monitored emission lines. Magenta and green points represent the PL emission intensity from the $X_{BE}^{0}$ and the $XX^{0}_{T_{\pm3}}$, as a function of the square root of the pulsed laser power tuned to the $s$-$p$ resonances of the BE and DE (color matched upward arrows in (c)). The color matched dashed lines present model fittings to the measured data. Intensities which correspond to $\pi$ and $2\pi$ area pulses are marked.}
\label{fig:PLE}
\end{center}
\end{figure}

The experimental study was performed on a sample grown by molecular-beam epitaxy on a (001)-oriented GaAs substrate. One layer of strain-induced InGaAs QDs was deposited in the center of a planar microcavity formed by two distributed Bragg reflecting mirrors. The microcavity was optimized for the range of wavelengths corresponding to photoluminescence (PL) emission caused by optical recombination between ground-state carriers in these QDs \cite{poem2007, kodriano2010,benny2012prb,benny2011prb}. The measurements were carried out at ~4.2 K. The experimental setup that we used provides spatial resolution of about 1~$\mu$m, spectral resolution of about 10 $\mu$eV, and temporal resolution of about 400 ps in measuring the arrival times of up to four different photons originating from up to four different spectral lines, where each line polarization can be projected on a different polarization direction.  More details about the sample \cite{garcִia1997,benny2011prb} and about the experimental setup \cite{kodriano2010, benny2011prb} are given in earlier publications. Pulsed excitation was performed using a frequency tunable, cavity-dumped dye laser synchronously pumped by a frequency-doubled Nd:YVO$_{4}$ (Spectra Physics Vanguard\texttrademark) laser. The spectral width was $\sim 150~\mu$eV, and the temporal width was about 10 ps, as reported in Ref. \cite{schwartz2015}. The repetition rate was set to 9.5 MHz, in which the cavity was dumping 1 out of 8 pulses. For temporally longer pulses, we used a grating-stabilized tunable diode laser modulated by an electro-optic modulator synchronized with the pulsed laser, permitting a variable pulse duration with rise and fall times of less than half a ns. The polarizations of the two lasers were independently controlled by a polarized beam splitter and two pairs of liquid crystal variable retarders (LCVRs). Similarly, the polarization of the detected PL was analyzed by an additional set of two pairs of LCVRs and a polarized beam splitter.

In Fig. \ref{fig:PLE}, we present measured PL and PLE spectra in which the various optical transitions discussed in Fig. \ref{fig:Scheme} are identified.
Fig. \ref{fig:PLE}(a,b) presents rectilinear polarization-sensitive PL spectra of the QD. The BE ($X^0_{BE}$), the DE ($X^0_{DE}$), and the positively-charged  (X$^{+1}$) excitonic lines as measured using very weak excitation by a 445 nm light source \cite{schwartz2015} as well as the XX$^0_{T\pm3}$ and XX$^0_{T0}$ (where the holes form a triplet with antiparallel spins \cite{poem2010,kodriano2010}) biexcitonic lines, measured using a considerably stronger excitation intensity, are clearly marked  in Fig. \ref{fig:PLE}(b) and Fig. \ref{fig:PLE}(a), respectively.
In Fig. \ref{fig:PLE}(c), we use PLE spectra to spectrally identify the excited BE and DE absorption resonances. To obtain the DE resonances (solid lines in Fig. \ref{fig:PLE}(c)), one laser was tuned to the XX$^0_{T\pm3}$ absorption resonance (upward blue arrow in Fig. \ref{fig:Scheme}(a)) while the other laser energy was scanned, and the emission from the XX$^0_{T\pm3}$ line (downward magenta arrow in Fig. \ref{fig:Scheme}(a)) was monitored. To obtain the BE resonances (dashed lines in Fig. \ref{fig:PLE}(c)), the emission from the BE was monitored during the same laser scan.

\begin{figure}[b]
\begin{center}
\includegraphics[width=\columnwidth]{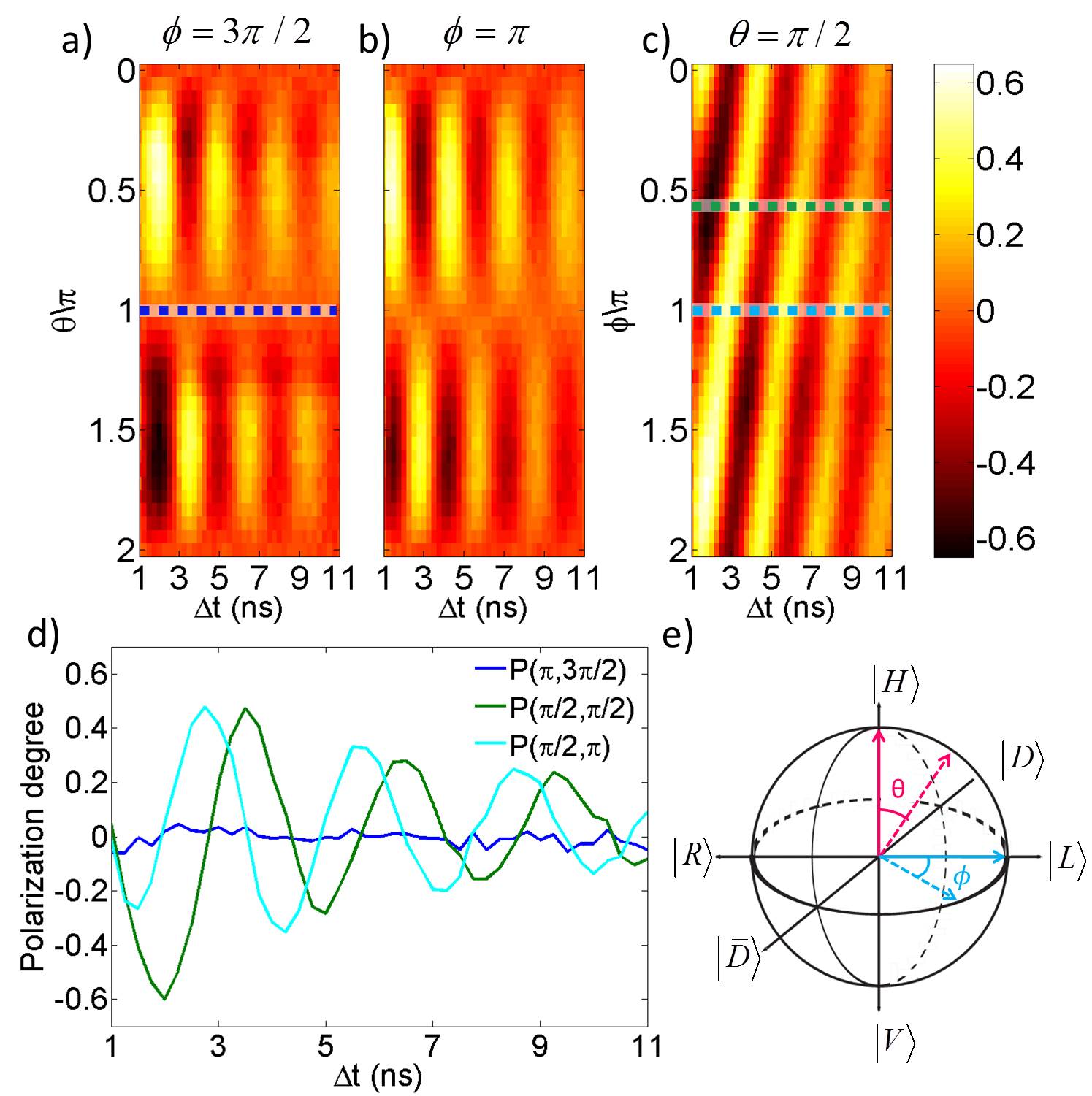}
\caption{(a) The degree of circular polarization of the  $XX^{0}_{T_{\pm3}}$ biexciton emission (represented by the color bar) as a function of time (measured from the writing laser pulse)  and the writing laser pulse polarization $\vec {P} (\theta,\phi)$. The degree of circular polarization was obtained by subtracting the PL resulting from left-hand circularly polarized probe pulse from that resulting from right-hand polarized probe pulse, dividing by their sum. The writing laser polarization was (a) $\vec {P} (\theta,3\pi/2)$, (b) $\vec {P} (\theta,\pi)$ and (c)$\vec {P} (\pi/2,\phi)$. (d) Polarization degree as a function of time for the pulse polarizations $\hat{V}=\vec {P} (\pi,0)$, $\hat {\bar {D}}=\vec {P} (\pi/2,\pi/2)$ and $\hat {R}=\vec {P} (\pi/2,\pi)$ polarizations. The colors of the curves match the colors of the horizontal lines on the images (a)-(c). (d) The writing pulse polarization is represented by the angles $\theta$ and $\phi$  on the  Poincar\'{e} sphere. }
\label{fig:Map}
\end{center}
\end{figure}

We identified the three lowest energy absorption resonances of the DE. One resonance at -0.3 meV is to its ground energy level ($(1e^1)_{\pm 1/2}(1h^1)_{\pm 3/2}$) and two others at 15 and 22 meV to its first ($(1e^1)_{\pm 1/2}(2h^1)_{\pm 3/2}$) and second ($(1e^1)_{\pm 1/2}(3h^1)_{\pm 3/2}$) excited levels, respectively. All three DE transitions are a few hundred $\mu eV$ below the corresponding BE transitions. We note that while the ground state resonance is horizontally linearly polarized \cite{schwartz2015,don2015}, the higher two transitions are unpolarized. Since excitation of the excited DE resonances does not result in any significant increase in the signal from the PLE of the BE, and vise versa, one safely concludes that during the relaxation (which we measured using pump probe techniques to be $\sim 0.1$ ns - not shown), spin flip processes are negligible. This perfectly agrees with previous experimental \cite{heiss2007, kodriano2010, benny2011prb} and theoretical \cite{woods2004,bulaev2005} studies.
In Fig. \ref{fig:PLE} (d), the PL emission intensity as a function of the square root of the average power of the pulsed laser tuned to the first excited resonances of the BE and DE (color matched to the arrows in (a), (b), and (c)) are presented by magenta and green points,  respectively. The dashed lines in Fig. \ref{fig:PLE} (d) present a model fitting to the experimental data.  Rabi oscillations~\cite{stievater2001, zrenner2002} are clearly observed for both the BE (magenta) and DE (green), and intensities which correspond to $\pi$ and $2\pi$ area pulses are marked.

The ratio between the oscillator strength of the first excited BE ($(1e^1)_{\pm 1/2}(2h^1)_{\mp 3/2}$) resonance to that of the DE ($(1e^1)_{\pm 1/2}(2h^1)_{\pm 3/2}$) is about 6.
The ratio between the oscillator strength of the ground BE state ($(1e^1)_{\pm 1/2}(1h^1)_{\mp 3/2}$) to that of the BE first excited resonance ($(1e^1)_{\pm 1/2}(2h^1)_{\mp 3/2}$) is about 35~\cite{benny2012prb}. Consequently, the oscillator strength of the excited DE resonance ($(1e^1)_{\pm 1/2}(2h^1)_{\pm 3/2}$) is about 200 times weaker than that of the ground state BE and about order of magnitude stronger than that of the ground state DE~\cite{schwartz2015}.
The increase in the DE oscillator strength, and its equal distribution between the two rectilinear polarizations, is attributed to increased DE-BE mixing~\cite{don2015} at these elevated energies. These properties of the excited DE  resonance enable deterministic photogeneration of the excited DE using few-ps long polarized optical pulses, while ``writing'' the DE spin state. The BE-DE mixing is small enough to prevent significant BE generation upon the spin-preserving phonon-assisted relaxation of the excited DE. The DE predominantly maintains its DE character~\cite{poem2010,kodriano2010}, as clearly demonstrated in Fig. \ref{fig:PLE}, allowing high-fidelity writing of the DE spin state.

The measured degree of circular polarization of the emission as a function of time after the writing pulse and the polarization of the writing pulse are presented in Fig. \ref{fig:Map}(a), \ref{fig:Map}(b) and \ref{fig:Map}(c) for $P(\theta,\phi)= P(0<\theta<2\pi, 3\pi/2)$, $P(0<\theta<2\pi,0)$ and $P(\pi/2,0< \phi <2\pi)$, respectively. The oscillations with a period of 3.1 ns correspond to the precession period of a coherent superposition of DE eigenstates and reflect the DE energy level splitting of 1.5 $\mu$eV. Changes in the polarization angle $\theta$ of the ``writing" pulse result in corresponding changes in the polarization visibility of the PL, while changes in the polarization angle $\phi$ of the writing pulse, result in corresponding changes in the polarization phase of the PL. Figs. \ref{fig:Map}(a)-\ref{fig:Map}(c) therefore demonstrate one-to-one correspondence between the initial amplitude and phase of the circular polarization oscillations  of the PL, which result from the precession of the DE and the angles $\theta$ and $\phi$, which describe the polarization of the writing pulse.

In Fig. \ref{fig:Map}(d), the temporal dependence of the degree of circular polarization is presented for various writing pulse polarizations. The selected polarizations are presented by the blue, green and azure lines, respectively. Thus, the pulse polarization deterministically ``writes" the dark exciton coherent spin state. The maximal degree of polarization that we obtain ($\sim$0.65) is compatible with the finite temporal resolution of our detectors ($\sim$400 ps) ~\cite{poem2010}, indicating above 90\% fidelity in writing the dark exciton qubit.

In summary, this work shows a one-to-one mapping between the polarization state of a picosecond long resonant laser pulse and the spin state of the quantum dot confined dark exciton that it deterministically photogenerates. The technique used in this work is similar to that used previously with the BE~\cite{benny2011prl}, though the dark exciton qubit lives three orders of magnitude longer and maintains it coherence for at least 100 ns after its generation~\cite{schwartz2015}. Combined with recent demonstrations of all optical depletion ~\cite{Schmidgall2015} of the QD from residual dark excitons and the ability to coherently control the dark exciton spin ~\cite{schwartz2015}, this work paves the way for exploiting the dark exciton as a particularly appealing matter spin qubit.
Moreover, there are many dark-exciton like metastable spin triplet states in matter. Many are known to have long lifetimes even at room temperature and to give rise to phosphorescence, rather than to photoluminescence which occurs in the optically allowed systems. The methods that we developed to coherently write and control the DE, should be applicable to those systems, as well.

\begin{acknowledgments}
The support of the Israeli Science Foundation (ISF),
the Technion's RBNI and the Israeli Nanotechnology Focal Technology Area on ``Nanophotonics for Detection" are gratefully acknowledged.
I. S. and D. C. contributed equally to this work.
\end{acknowledgments}

\bibliographystyle{unsrt}
\bibliography{write_with_phase_bib}

\end{document}